\documentclass[aps,twocolumn,showpacs,floatfix,preprintnumbers,amsmath,amssymb]{revtex4}
\usepackage{epsfig}  % Include figure files
\begin{document}

\title{Excited decuplet baryons from QCD sum rules}
\author{Frank X. Lee}
\affiliation{Center for Nuclear Studies, Department of Physics,
The George Washington University,  Washington, DC 20052, USA}

\begin{abstract}
A calculation of the mass spectrum in the baryon decuplet sector is presented using the 
method of QCD sum rules. Sum rules are derived for states of 
spin-parity $3/2\pm$ and $1/2+$ using both the conventional method and a parity-projection method.
The predictive ability of the sum rules is explored by a Monte-Carlo based
analysis procedure in which the three phenomenological parameters
(mass, coupling, threshold) are treated as free parameters and
fitted simultaneously. Taken together, the results give an improved determination of the 
mass spectrum in this sector from the perspective of non-perturbative QCD.
\end{abstract}
\vspace{1cm}
 % Monte Carlo and statistical methods
%\keywords{QCD sum rule method, baryon spectroscopy}
\maketitle

%%%%%-------------------------------------------------------------------
\section{Introduction}
\label{intro}

A goal of hadronic physics is to understand the baryon spectrum from QCD, the underlying
theory of the strong interaction. 
Experimentally, the push is fueled by the physics program at JLab, and other accelerator facilities. 
To this end, the time-honored QCD sum rule method~\cite{SVZ79} adds a useful theoretical tool.
It is a non-perturbative approach to QCD that reveals a direct connection between
hadronic observables and the QCD vacuum structure via a few universal parameters
called vacuum condensates (vacuum expectation values of QCD local operators),
the most important of which are
the quark condensate, the mixed condensate and the gluon condensate.
The method is analytical, physically transparent, and
has minimal model dependence with well-understood
limitations inherent in the operator product expansion (OPE).
It provides a complementary view of the same non-perturbative physics 
to the numerical approach of lattice QCD.
The method was applied to the decuplet sector not long after it was 
introduced~\cite{Bely82,Bely83,RRY82,RRY85,Chung84}. However, only limited 
attention has been paid to this sector since then. 
In Ref.~\cite{Derek90}, a systematic study was made that incudes some $\Delta$ states.
Later, progress came in the analysis of QCD sum rules by Monte-Carlo sampling of errors~\cite{Derek96}.
It was applied to the study of spin-3/2 branch of baryon decuplet~\cite{Lee98}, magnetic moments~\cite{Lee98b,Lee98c} 
and other spin-3/2 excited baryons~\cite{Lee02}.

From a theoretical standpoint, 
the main issue is how to isolate sum rules that couple to a particular spin-parity,
since the interpolating fields used to construct the spin-3/2 states 
contain both spin-3/2 and spin-1/2 components, and couple to both parities.
Separation of the spin-3/2 and spin-1/2 components can be achieved by examining the 
Dirac structures upon which the sum rules are based, as done in Ref.~\cite{Derek90}.
However, the parities are still mixed. In the pole-plus-continuum ansatz of the method, 
isolation of a particular parity relies on strong cancellations in the excited part of the spectrum.
Without such cancellations, the sum rules usually suffer contaminations from higher states,
 a problem that renders them unstable.  
A solution to separate the parities exactly was proposed in Ref.~\cite{Jido97} which 
showed improved convergence in the octet baryon sector.
In this work, we apply the method to the decuplet sector, together with the conventional method.
The goal is to identify sum rules that couple strongly to the four branches 
of the decuplet with spin-parity $3/2+$, $3/2-$, $1/2+$, and $1/2-$.
In addition to the Delta states, 
we also examine states that contain the strange quark, 
which have not been studied in detail.
In our calculation, we consistently include operators up to dimension eight,
first order strange quark mass corrections,
flavor symmetry breaking of the strange quark condensates,
anomalous dimension corrections, and possible factorization violation of
the four-quark condensate. Furthermore, we try to assess
quantitatively the errors in the phenomenological parameters,
using the Monte-Carlo based analysis procedure. This
procedure incorporates all uncertainties in the QCD input
parameters simultaneously, and translates them into uncertainties
in the phenomenological parameters, with careful regard to OPE
convergence and ground state dominance.

%%%%%-------------------------------------------------------------------
\section{Method}
\label{elem}

In the conventional method, 
the starting point is the time-ordered two-point
correlation function in the QCD vacuum
\begin{equation}
\Pi(p)=i\int d^4x\; e^{ip\cdot x}\,\langle 0\,|\,
T\{\;\eta(x)\, \bar{\eta}(0)\;\}\,|\,0\rangle. \label{cf2pt-old}
\end{equation}
where $\eta$ is the interpolating field with
the quantum numbers of the baryon under consideration. 
Assuming SU(2) isospin symmetry in the u and d quarks, we consider
the most general current of spin 3/2 and isospin 3/2 for the $\Delta$,
\begin{equation}
\eta^{\Delta}_{\mu}(x)=\epsilon^{abc}[
u^{aT}(x)C\sigma_{\mu} u^b(x)]u^c(x)
\end{equation}
Here $C$ is the charge conjugation operator and the superscript
$T$ means transpose. The antisymmetric $\epsilon^{abc}$ and sum over color ensures 
a color-singlet state.
For the other members of the decuplet, we consider, omitting the explicit $x$-dependence
\begin{equation}
\eta^{\Sigma^*}_{\mu}=\sqrt{1/3}\epsilon^{abc}\{2[
u^{aT}C\sigma_{\mu} s^b]u^c
+ [u^{aT}C\sigma_{\mu} u^b]s^c \},
\end{equation}
\begin{equation}
\eta^{\Xi^*}_{\mu}=\sqrt{1/3}\epsilon^{abc}\{2[
s^{aT}C\sigma_{\mu} u^b]s^c
+ [s^{aT}C\sigma_{\mu} s^b]u^c \},
\end{equation}
\begin{equation}
\eta^{\Omega^-}_{\mu}=\epsilon^{abc}[
s^{aT}C\sigma_{\mu} s^b]s^c.
\end{equation}

A baryon interpolating field couples to both the ground state and the
excited states of a baryon, and to both parities. That leads to 
sum rules that contain mixed parity states. 
The parities can be separated by considering the `forward-propagating' version of the 
correlation function~\cite{Jido97},
\begin{equation}
\Pi(p)=i\int d^4x\; e^{ip\cdot x}\,\theta(x_0)\langle 0\,|\,
T\{\;\eta(x)\, \bar{\eta}(0)\;\}\,|\,0\rangle. \label{cf2pt}
\end{equation}
where the only difference between the above equation and the conventional
two-point correlation function in Eq.~(\ref{cf2pt-old}) is the step function $\theta(x_0)$.
Under this condition, the phenomenological representation of the 
the imaginary part in the rest frame ($\vec{p}=0$) can be written as 
\begin{eqnarray}
\mbox{Im}\,\Pi(p_0) &=& 
\sum_n \left[ (\lambda_n^+)^2 \frac{\gamma_0+1}{2}\delta(p_0-m_n^+)
\nonumber \right. \\ 
&+& \left. (\lambda_n^-)^2 \frac{\gamma_0-1}{2}\,\delta(p_0-m_n^-) \right] 
\nonumber \\ 
&\equiv& \gamma_0 A(p_0) + B(p_0)
\label{Img_Pi}
\end{eqnarray}
where $\lambda^\pm$ and $m^\pm$ are the couplings and masses of the states involved, 
separated by parity.
The functions in the last step are defined by
\begin{equation}
A(p_0) =\, \frac{1}{2}\sum_n\,[(\lambda_n^+)^2
\delta(p_0-m_n^+)\,+\,(\lambda_n^-)^2\delta(p_0-m_n^-)],
\label{A}
\end{equation}
and 
\begin{equation}
B(p_0) =\, \frac{1}{2}\sum_n\,[(\lambda_n^+)^2
\delta(p_0-m_n^+)\,-\,(\lambda_n^-)^2\delta(p_0-m_n^-)]. 
\label{B}
\end{equation}
By considering the combinations $A(p_0)+B(p_0)$ and $A(p_0)-B(p_0)$,
positive-parity and negative-parity states are separated exactly, respectively.
This method is reminiscent of the one used in lattice QCD under Dirichlet 
boundary conditions in the time direction~\cite{Lee06}.

The construction of sum rules in Borel space proceeds by 
taking the integral with a weighting factor and 
truncating all excited states starting at a certain threshold.
For positive-parity states,
\begin{equation}
 \int^{w_+}_0 \left[ 
A(p_0)+B(p_0) \right] e^{-p_0^2/M^2}\,dp_0
= \lambda_+^2e^{-m_+^2/M^2}
\label{G_sum_rule1}
\end{equation}
and for positive-parity states,
\begin{equation}
 \int^{w_-}_0 \left[ 
A(p_0)-B(p_0) \right] e^{-p_0^2/M^2}\,dp_0
= \lambda_-^2e^{-m_-^2/M^2}.
\label{G_sum_rule2}
\end{equation}

To separate out the spin components, we need the full Dirac structure of the
correlation function for spin-3/2 interpolating fields~\cite{Derek90,Lee98}, 
\begin{eqnarray}
\Pi_{\mu\nu}(p) &=& 
\lambda^2_{3/2}\; \left\{ -g_{\mu\nu}\hat{p}
+{1\over 3}\gamma_\mu \gamma_\nu \hat{p}
\right. \nonumber \\ &-& \left. 
 {1\over 3}(\gamma_\mu p_\nu-\gamma_\nu p_\mu)
+ {2\over 3}{p_\mu p_\nu \over M^2_{3/2}}
\right. \nonumber \\ &\pm& \left. 
    M_{3/2} \left[ g_{\mu\nu} 
- {1\over 3}\gamma_\mu\gamma_\nu
\right. \right. \nonumber \\ &+& \left. \left.
 {1\over 3 M^2_{3/2}}(\gamma_\mu p_\nu-\gamma_\nu p_\mu)\hat{p}
- {2\over 3}{p_\mu p_\nu \over M^2_{3/2}} \right]
\right\} \nonumber \\ &+&
  \alpha^2_{1/2}\; \left\{ {16\over M^2_{1/2}} p_\mu p_\nu\hat{p}
-\gamma_\mu \gamma_\nu \hat{p}
\right. \nonumber \\ &-& \left. 
 6{ \over }(\gamma_\mu p_\nu+\gamma_\nu p_\mu)
+4(\gamma_\mu p_\nu-\gamma_\nu p_\mu)
\right. \nonumber \\ &\pm & \left. 
 M_{1/2} \left[ -\gamma_\mu\gamma_\nu
- {8 p_\mu p_\nu \over M^2_{1/2}}
\right. \right. \nonumber \\ &+& \left. \left.
 {4\over M^2_{1/2}}(\gamma_\mu p_\nu-\gamma_\nu p_\mu)\hat{p}
\right]\,\right\}
+ \cdots,
\label{correlator33}
\end{eqnarray}
where the ellipses denote excited state contributions, 
and $\hat{p}$ denotes $p_\mu \gamma^\mu$.
The upper/lower sign corresponds to the explicit positive/negative parity when a state 
contributes.
The entire structure is divided into two parts: one for spin-3/2 with coupling $\lambda^2_{3/2}$,
and the other spin-1/2 with coupling $\alpha^2_{1/2}$. 
The structures are chosen to be effectively orthogonal to each other. 
Only six of the structures are independent.
The two sum rules from $g_{\mu\nu}\hat{p}$ and $g_{\mu\nu}$ 
couple only to spin-3/2 states because they do not appear in the spin-1/2 part.
The two sum rules from $\gamma_\mu p_\nu+\gamma_\nu p_\mu$ and 
$\gamma_\mu \gamma_\nu+\frac{1}{3}g_{\mu\nu}$ couple only to spin-1/2 states because 
they do not appear in the spin-3/2 part.
By the same token, the sum rule at $\gamma_\mu p_\nu+\gamma_\nu p_\mu$ that couples to spin-1/2 
could also be obtained from the combination 
%$(\gamma_\mu p_\nu-\gamma_\nu p_\mu)-\frac{1}{3}g_{\mu\nu}\hat{p}$ 
$\gamma_\mu \gamma_\nu\hat{p}+\frac{1}{3}g_{\mu\nu}\hat{p}$.
The two sum rules from $p_\mu p_\nu\hat{p}$ and $p_\mu p_\nu$
couple to both spin-3/2 and spin-1/2 states because they appear in both parts.
They will not be considered further.
Note that the parities are still mixed in each of these sum rules.
It is not known {\it a priori} which parity is the dominant state in a given sum rule.
That information comes from the specific OPE structure on the left-hand-side.

%%%%%-------------------------------------------------------------------
\section{Parity-projected QCD sum rules for spin-3/2 decuplet states}
\label{qcd33}

The combination of the two tensor structures that couple purely to spin-3/2 states, 
$g_{\mu\nu}\hat{p}$ and $g_{\mu\nu}$, can be cast into the form 
$g_{\mu\nu}(\gamma_0 A + B)$ in the rest frame.
According to Eq.~(\ref{Img_Pi}), it suggests that parity projection can be 
done exactly in this case.
The functions $A(p_0)$ and $B(p_0)$ are readily identified from the usual 
calculation of the OPE which involves contracting out the quark pairs 
in the correlation function and substituting the fully-interacting quark propagator 
from OPE.
They are functions of the QCD vacuum condensates and other QCD parameters.

Upon Borel integration in Eq.~(\ref{G_sum_rule1}) and Eq.~(\ref{G_sum_rule2}),
the parity-projected sum rules can be written in the general form 
\begin{equation}
{\cal A}(w_+,M)\,+\,{\cal B}(w_+,M)\, =\,
\tilde{\lambda}_{+}^2 e^{-m_{+}^2/M^2}, 
\label{F_sum_rule1}
\end{equation}
\begin{equation}
{\cal A}(w_-,M)\,-\,{\cal B}(w_-,M)\, =\,
\tilde{\lambda}_{-}^2 e^{-m_{-}^2/M^2}, 
\label{F_sum_rule2}
\end{equation}
where $\tilde{\lambda}_{\pm}=(2\pi)^2\lambda_{\pm}$ are the rescaled couplings. 
The rescaling is done so that no factors of $\pi$ appear explicitly in the sum rules.
The rescaled quantities are also `natural' in their numerical values.

For our special case of spin-3/2 states, the ${\cal A}$ function ${\cal A}_{3/2}$ is given by
\begin{eqnarray}
 {\cal A}_{3/2} &=& 
c_1\,[1]\,(2(1-e^{-w^2/M^2})M^6 - 2w^2e^{-w^2/M^2}M^4 
\nonumber \\ 
&-& w^4e^{-w^2/M^2}M^2)\,L^{4/27} 
\nonumber \\
&+ & (c_2\,[b] +c_3\,[m_sa])\,(1-e^{-w^2/M^2})M^2 L^{4/27}\, 
\nonumber \\
&+& c_4 \,[m_s m_0^2 a]\, L^{-10/27} +\,c_5 \, [\kappa_v a^2]\, L^{28/27}. 
\label{sum33A}
\end{eqnarray}
In this expression, the vacuum condensates are explicitly isolated in square brackets whose 
definitions and values will be given below.
The Wilson coefficients (which are pure numbers) are given for various members of the decuplet by
\begin{equation}
\begin{array}{|c|c|c|c|c|c|c|c}
\hline
{\cal A}_{3/2}&c_1    &c_2           &c_3              &c_4                 &c_5 \\
\hline
\Delta   &{1\over 20} &{-5\over 144} &0                &0                   &{-2\over 3} \\
\hline
\Sigma^* &{1\over 20} &{-5\over 144} &{(4-f_s)\over 6} &{-(14-5f_s)\over 36}&{(2+4f_s)\over 9} \\
\hline
\Xi^*    &{1\over 20} &{-5\over 144} &{(2+f_s)\over 3} &{-(7+2f_s)\over 18} &{2f_s(2+f_s)\over 9} \\
\hline
\Omega^-   &{1\over 20} &{-5\over 144} &{3f_s\over 2}    &{-3f_s\over 4}      &{-2f_s^2\over 3}.\\
\hline
\end{array}
\end{equation}
Similarly, the ${\cal B}_{3/2}$ function is given by
\begin{eqnarray}
 {\cal B}_{3/2} &=& 
c_1\,[m_s](I(w)-we^{-w^2/M^2})L^{-8/27}M^4\, 
\nonumber \\
&+& c_2 [m_s] w^3\,e^{-w^2/M^2}L^{-8/27}M^2 
\nonumber \\
&+& c_3\,[a] (I(w) - we^{-w^2/M^2})L^{16/27}M^2
\nonumber \\
&+& c_4\,[m_s b] I(w) L^{-8/27} 
\nonumber \\
&+& c_5\,[m_0^2 a] I(w) L^{2/27},
\label{sum33B}
\end{eqnarray}
where the coefficients are given  by
\begin{equation}
\begin{array}{|c|c|c|c|c|c|}
\hline
{\cal B}_{3/2}&c_1    &c_2           &c_3              &c_4                 &c_5 \\
\hline
\Delta   & 0    & 0     & {2\over 3}         & 0     & {-2\over 3}      \\
\hline
\Sigma^* & 3/16 & -1/8  & {2(2+f_s)\over 9}  & -1/24 & {-2(2+f_s)\over 9}  \\
\hline
\Xi^*    & 3/8  & -1/4  & {2(1+2f_s)\over 9} & -1/12 & {-2(1+2f_s)\over 9}  \\
\hline
\Omega^-   & 9/16 & -3/8  & {2f_s\over 3}      & -1/8  & {-2f_s\over 3}.      \\
\hline
\end{array}
\end{equation}
The integral $I(w)=\int_0^w e^{-x^2/M^2} dx$ is evaluated numerically in the analysis.
Note that the sum rule for $\Omega^-$ reduces to that for the $\Delta$ if $m_s=0$ and $f_s=1$, 
which serves as a check of the calculation.

Now we explain the meaning of the parameters in the sum rules.
The rescaled quark condensate is taken as the positive quantity
$a=-(2\pi)^2\,\langle\bar{u}u\rangle=0.52\pm 0.05$ GeV$^3$,
corresponding to a central value of $\langle\bar{u}u\rangle=-(236)^3$ MeV$^3$.
For the gluon condensate, we use rescaled $b=\langle g^2_c\, G^2\rangle=1.2\pm 0.6$ GeV$^4$.
The mixed condensate parameter is placed at
$m^2_0=\langle\overline{q}g\sigma\cdot Gq\rangle/\langle\overline{q}q\rangle=0.72\pm 0.08$ GeV$^2$.
For the four-quark condensate $\langle\bar{u}u\bar{u}u\rangle=\kappa_v\langle\bar{u}u\rangle^2$,
we use $\kappa_v=2\pm 1$ to include possible violation of the factorization approximation.
We retain only terms linear in the strange quark mass and set $m_u=m_d=0$.
The strange quark mass is taken as $m_s=0.15\pm 0.02$ GeV.
We use the ratio $f_s=\langle\bar{s}s\rangle/\langle\bar{u}u\rangle
=\langle\bar{s}g_c\sigma\cdot G s\rangle /\langle\bar{u}g_c\sigma\cdot G u\rangle
=0.83\pm 0.05$ to accounts for the flavor symmetry breaking of the strange quark condensates.
The anomalous dimension corrections of the various
operators are taken into account via the factors
$L^\gamma=\left[{\alpha_s(\mu^2)/ \alpha_s(M^2)}\right]^\gamma
=\left[{\ln(M^2/\Lambda_{QCD}^2)/
\ln(\mu^2/\Lambda_{QCD}^2)}\right]^\gamma$, where $\gamma$ is the
appropriate anomalous dimension, $\mu=500$ MeV is the
renormalization scale, and $\Lambda_{QCD}=0.15\pm 0.04$ GeV is the QCD scale parameter.
We find variations of $\Lambda_{QCD}$ have little effects on the results.
The uncertainties are assigned fairly conservatively, ranging from 10\% to 100\%. 
They will be mapped into those for the fit parameters in the analysis, giving a realistic estimate 
of the errors on the parameters.

It is worth pointing out that the $A$ function is chiral-even, meaning that 
it only involves condensates of even energy dimensions:
for example, $b$ of 4, $m_sa$ of 4, $\kappa_va^2$ of 6 and so on.
On the other hand, the $B$ function is chiral-odd, with the leading contribution from the 
quark condensate $a$, the order parameter of spontaneous chiral symmetry breaking of QCD.  
The difference of the two parity-projected sum rules 
in Eq.~(\ref{F_sum_rule1}) and Eq.~(\ref{F_sum_rule2}) is the sign in front of the
chiral-odd term $B$. So from the point of view of QCD sum rules, the origin of splittings 
between positive-parity and negative-parity states lies in the chiral-odd vacuum condensates.
If the quark condensate and the strange quark mass vanish, 
then chiral symmetry is restored in the vacuum, and 
there would be exact parity doubling in the baryon spectrum.
This is a valuable insight from the parity-projected QCD sum rules.

The analysis of a QCD sum rule boils down to the following mathematical problem.
Given an equation of the general structure
\begin{equation}
\mbox{LHS}(M,OPE)=\mbox{RHS}(M,w,m,\lambda^2),
\label{sum}
\end{equation}
and a set of input QCD parameters denoted by $OPE$, find the best output parameters
(the baryon mass of interest $m$, the coupling strength $\lambda^2$ of the interpolating field, 
and the continuum threshold $w$) by matching the two sides over a region in Borel mass $M$.
The LHS has errors arising from our imprecise knowledge of the QCD parameters, as discussed above.
From statistical point of view, a $\chi^2$ minimization of the type
\begin{equation}
\chi^2=\sum_i{|\mbox{LHS}_i-\mbox{RHS}_i|^2 \over \sigma^2_i},
\label{chi2}
\end{equation}
offers the least-biased way of finding the unknown parameters.
In the Monte-Carlo based procedure used here,
first the Borel window in $M$ is divided equally into a grid (we use 51 points).
At each point, the uncertainty distribution in the OPE is constructed
by randomly-selected, Gaussianly-distributed sets generated
from the central values and the uncertainties in the QCD input parameters.
Then the $\chi^2$ minimization is applied to the
sum rule by fitting the phenomenological parameters. This is
done for each QCD parameter set and at each point in $M$, resulting in distributions for
phenomenological fit parameters, from which their errors are derived.
Usually, 100 such configurations are sufficient for getting stable
results. We generally select 500 sets which help resolve more
subtle correlations among the QCD parameters and the phenomenological fit parameters.
So the same sum rule is fitted not just once, but thousands of times in our analysis.
The advantage afforded by the Monte-Carlo is that the entire phase-space of the
input QCD parameters is explored. The errors obtained this way represent
the most realistic and conservative estimates of the predictive power of a QCD sum rule.
This is in contrast to traditional approaches
where only a small part of the phase-space is explored at a time.
We select the Borel window by trial and error. Our criteria are that
the OPE is reasonably convergent by looking at the terms of various dimension,
and that the fit results should not be sensitive to small changes in the Borel window.
Under these constraints, we seek to make the window as wide as possible.
Since the selection of Borel window $M$ and the three phenomenological parameters
are inter-dependent in the sum rule,
the entire fitting process is iterated until the best solution is found.
In general, we seek solutions that are from a three-parameter search
in which  all three parameters are treated as free simultaneously.
We regard such solutions as the best predictions of the QCD sum rule approach.
In the absence of such solutions, a two-parameter search is performed in which one of the
three parameters is fixed. Usually, it is the continuum threshold which is fixed to values
that are larger than the mass in question, or to values suggested by the observed spectrum.
Such a two-parameter approach is the one usually adopted in most analyses in the past.

\begin{figure*}  % * means big figure across the page
\centerline{\psfig{file=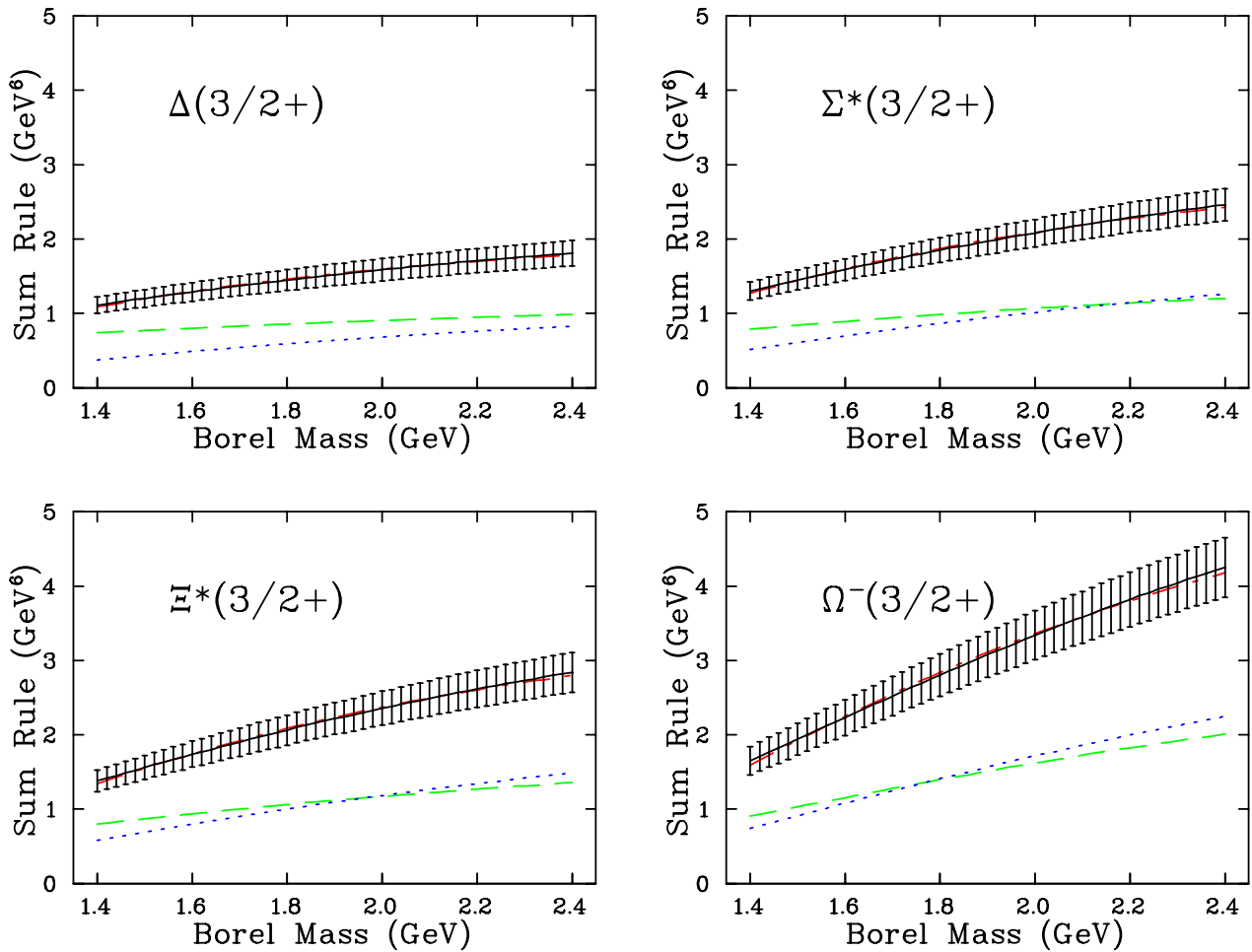,width=5.0in,angle=90}}
\caption{Matching of the sum rules in Eq.~(\protect\ref{F_sum_rule1})
for positive-parity spin-3/2 states as a function of the Borel mass.
The solid line along with the error band generated via Monte-Carlo is the LHS $({\cal A}+{\cal B})$. 
The dash-dot line, which is barely distinguishable from the solid line, 
is the RHS (ground-state pole).
Also plotted are the individual contributions of term ${\cal A}$ (dotted line) and 
term ${\cal B}$ (dashed line).}
\label{fit1}
\end{figure*}

Fig.~\ref{fit1} shows the matching in the sum rule Eq.~(\ref{F_sum_rule1}) for $3/2+$ states.
The two sides match very well over a relatively wide region (1 GeV).
The individual contributions of term ${\cal A}$ and term ${\cal B}$ are also shown.
The two terms are comparable in size, the sum of which gives rise to the LHS.
The error band on the LHS is generated by Monte-Carlo reflecting all the uncertainties 
assigned to the QCD parameters. The entire uncertainty phase space is then mapped 
to the output parameters in the fitting process.
%
%%%%%%%%%%%%%%%%%%%%%%%%%%%%%%%%%%%%%%%%%%%%%%%%%%%%%
\begin{table}
\caption{Results for the $3/2+$ states from the parity-projected sum rule
Eq.(~\protect\ref{F_sum_rule1}) in a two-parameter search. 
The errors are derived from 500 Monte-Carlo sets based on the uncertainties assigned 
to the QCD input parameters. Two sets of solutions are provided for each case: 
the first one using the default errors in the OPE input parameters, 
while the second using uniform 10\% errors.
The experimental values are taken from the PDG~\protect\cite{pdg04}.}
\label{tab3p}
\begin{tabular}{l|ccccl}
\hline
      & Region & $w$ & $\tilde{\lambda}_{3/2+}^2$ & Mass & Exp.\\
      & (GeV) &  (GeV) & (GeV$^6$) & (GeV) & (GeV)\\ \hline
$\Delta({3\over 2}+)$
& 1.4 to 2.4 & 1.60 & 2.46$\pm$0.42 & 1.21$\pm$0.06 & 1.23 \\
&            &      & 2.32$\pm$0.21 & 1.21$\pm$0.02 &       
\\
\hline
$\Sigma^*({3\over 2}+)$
& 1.4 to 2.4  & 1.84 & 3.52$\pm$0.43 & 1.38$\pm$0.06 & 1.385 \\
&             &      & 3.40$\pm$0.29 & 1.37$\pm$0.02 &      
\\
\hline
$\Xi^*({3\over 2}+)$
& 1.4 to 2.4 &  1.95  & 4.19$\pm$0.46 & 1.47$\pm$0.05 & 1.53 \\
&            &        & 4.10$\pm$0.37 & 1.47$\pm$0.02 &      
\\
\hline
$\Omega^-({3\over 2}+)$
& 1.4 to 2.4 &  2.25  &6.94$\pm$0.63 & 1.68$\pm$0.05 & 1.67 \\
&            &        &6.89$\pm$0.56 & 1.68$\pm$0.02 &     
\\ \hline
\end{tabular}
\end{table}
%
%%%%%%%%%%%%%%%%%%%%%%%%%%%%%%%%%%%%%%%%%%%%%%%%%%%%%
%
A three-parameter search was tried first. Unfortunately, a solution could not be found.
What happens is that the search algorithm 
keeps returning solutions 
with threshold smaller than the mass, a clearly unphysical situation.
This is an indication that the OPE does not have enough information to resolve all three 
parameters simultaneously. 
So we switched to a two-parameter search by fixing the continuum 
threshold to values suggested by the Particle Data Group~\cite{pdg04}.
The extracted parameters are given in Table~\ref{tab3p}.
We offer two sets of results on the fit parameters: one using the default errors assigned to the 
OPE parameters ranging from 10\% to 100\%, the other with 10\% uniform errors.
It sort of gives the worst-case, best-case scenarios, as far as errors on the input parameters are concerned.
In the worst-case solution, the errors on the masses are on the order of 5\%, and 15\% on the couplings.
In the best-case solution, the errors are reduced to  about 2\% on the masses, 
and less than 10\% on the couplings.
Note that the central value for the couplings are shifted by a small amount 
when the errors are reduced, while it is stable for the masses.
The computed masses compare favorably with experiment. 
What is interesting is the fact that the mass pattern emerges under the same Borel 
window across the particles, and that the continuum thresholds are consistent with the 
excited states suggested by PDG.
In the case of $\Xi^*(3/+)$, the computed mass of 1.47 GeV 
is slightly below the experimental value of 1.53 GeV.
In the PDG, there are two 3-star excited states at 
1690 MeV and 1950 MeV with unknown spin-parity.
Our sum rule favors a threshold of 1950 MeV to 1690 MeV which leads to a mass that is 
even smaller than 1.47 GeV. It hints that the spin-parity of 1690 MeV is 
likely not $3/2+$.

\begin{figure*}
\centerline{\psfig{file=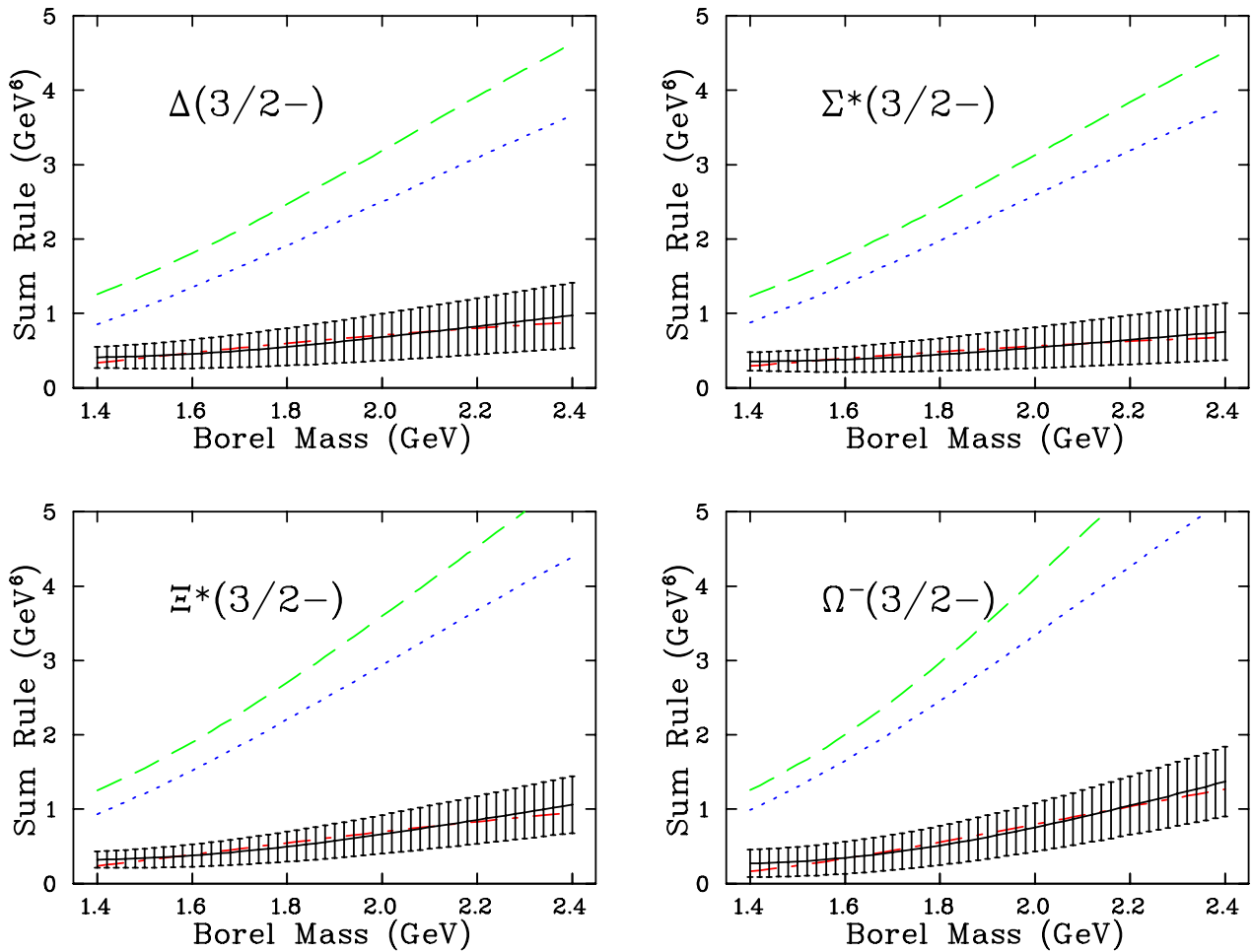,width=5.0in,angle=90}}
\caption{Matching of the sum rules in Eq.~(\protect\ref{F_sum_rule2})
for negative-parity spin-3/2 states.
The solid line along with the error band generated via Monte-Carlo is the LHS $({\cal A}-{\cal B})$. 
The dash-dot line, which is barely distinguishable from the solid line, 
is the RHS (ground-state pole).
Also plotted are the individual contributions of term ${\cal A}$ (dotted line) and 
term ${\cal B}$ (dashed line).}
\label{fit2}
\end{figure*}

Fig.~\ref{fit2} shows the matching in the sum rule Eq.~(\ref{F_sum_rule2}) for $3/2-$ states.
Here it is the difference between term ${\cal A}$ and term ${\cal B}$ that gives rise to the 
ground-state pole. The quality of the matching is still fairly good, 
but we found that the stability of the sum rules 
is not as good as the corresponding ones for $3/2+$ states in Eq.~(\ref{F_sum_rule1}).
A three-parameter search does not work, so a two-parameter search is performed.
To get a better understanding of the sum rules, we adopt the following fitting strategy. 
Instead of fixing the continuum threshold to a value that leads to a known state, 
we perform two-parameter searches by fixing the mass to values suggested by the PDG,
and search for the continuum threshold and coupling. 
In this way, the searches will provide information on whether a specific sum rule can 
accommodate a known state with reasonable continuum thresholds and couplings. 
Such a study is useful since to our knowledge 
there is no information about these $3/2-$ states from the standpoint of QCD sum rules.
The result of the study is given in Table~\protect\ref{tab3n}.
We fix the Borel window to be the same wide window (1.4 to 2.4 GeV) 
as for the $3/2+$ states for all cases. This choice is
quite reasonable judging by the matching plots, 
which are all of comparable quality as Fig.~\ref{fit2}.
Again, the same worst-case, best-case solutions are provided for each case in order to 
give some idea about the stability of the fits.

In the case of $\Delta({3/2}-)$, the PDG lists a 4-star state at 1.70 GeV.
Using this value as input, a continuum threshold of 2.74 GeV and a coupling of 1.56 
are obtained, with relative errors about 20\% and 55\%, respectively.
They decrease to about 5\% and 40\% in the best-case (10\% uniform errors on the input parameters).
Note that the central values vary as errors are reduced: 
the threshold shifts up to 2.9 GeV and the coupling shifts down to 1.43.
It is a sign of highly non-linear mapping of the errors from input to output.
These errors are much bigger than those for $3/2+$ states, especially on the couplings.
The next $\Delta(3/2-)$ state in the PDG is a one-star state at 1.94 GeV, compared to 
the continuum threshold of 2.7 to 2.9 GeV from this sum rule.

In the case of $\Sigma^*({3/2}-)$, the PDG lists a 2-star state at 1.58 GeV with unknown spin-parity, 
a 4-star state at 1.67 GeV, and a three-tar state at 1.94 GeV.
The corresponding continuum thresholds required to predict these states are 2.87, 2.90, 3.00 GeV. 

In the case of $\Xi^*({3/2}-)$, out of the three 3-star states listed by the PDG,
1.69, 1.82, 1.95 GeV, only the 1.82 GeV state is assigned the $3/2-$ spin-parity.
The corresponding continuum thresholds required to predict these states are 2.91, 2.95, 3.00 GeV. 
Interestingly, these values are close to the ones for the $\Sigma^*({3/2}-)$ states.

In the case of $\Omega^-({3/2}-)$, there are three candidates from the PDG, 
a three-star state at 2.25 GeV, a two-star state at 2.38 GeV, and a two-star state at 2.47 GeV, 
all with unknown spin-parity. Using the sum rules, they could be assigned $3/2-$ if the 
the corresponding continuum thresholds are 3.11, 3.18, 3.24 GeV. 
For the 2.25 GeV state, we could find a solution only when the errors are small.

%%%%%%%%%%%%%%%%%%%%%%%%%%%%%%%%%%%%%%%%%%%%%%%%%%%%%
%
\begin{table}
\caption{Results for the $3/2-$ states from the parity-projected sum rule
Eq.(~\protect\ref{F_sum_rule2}) in a two-parameter search. 
The errors are derived from 500 Monte-Carlo sets based on the uncertainties assigned 
to the QCD input parameters. Two sets of solutions are provided for each case: 
the first one using the default errors in the OPE input parameters, 
while the second using uniform 10\% errors.
The experimental values are taken from the PDG~\protect\cite{pdg04}.}
\label{tab3n}
\begin{tabular}{c|ccccl}
\hline 
          & Region & $w$ & $\tilde{\lambda}_{3/2-}^2$ & Mass & Exp.\\
          & (GeV) &  (GeV) & (GeV$^6$) & (GeV) & (GeV)\\ \hline
$\Delta({3\over 2}-)$
& 1.4 to 2.4 & 2.74$\pm$0.47 & 1.56$\pm$0.84 & 1.70 &  1.70 \\
&            & 2.90$\pm$0.10 & 1.43$\pm$0.56 &      &      
\\
\hline 
$\Sigma^*({3\over 2}-)$
& 1.4 to 2.4  & 2.68$\pm$0.51 & 1.15$\pm$0.64 & 1.58 & 1.58(?) \\
&             & 2.87$\pm$0.13 & 1.02$\pm$0.39 &      &     
\\
&             & 2.74$\pm$0.46 & 1.31$\pm$0.72 & 1.67 & 1.67 \\
&             & 2.90$\pm$0.11 & 1.19$\pm$0.46 &      &     
\\
&             & 2.94$\pm$0.30 & 2.14$\pm$1.20 & 1.94 & 1.94 \\
&             & 3.00$\pm$0.06 & 1.94$\pm$0.76 &      &     
\\
\hline 
$\Xi^*({3\over 2}-)$
& 1.4 to 2.4 &  2.72$\pm$0.43 & 1.13$\pm$0.65 & 1.69 & 1.69(?) \\
&            &  2.91$\pm$0.12 & 1.02$\pm$0.42 &      &        
\\
&            &  2.86$\pm$0.36 & 1.40$\pm$0.81 & 1.82 & 1.82 \\
&            &  2.95$\pm$0.10 & 1.28$\pm$0.53 &      &     
\\
&            &  2.94$\pm$0.30 & 1.80$\pm$1.05 & 1.95 & 1.95(?) \\
&            &  3.00$\pm$0.10 & 1.63$\pm$0.68 &      &        
\\
%&            &  2.99$\pm$0.27 & 2.11$\pm$1.25 & 2.03 & 2.03(?) \\
%&            &  3.03$\pm$0.10 & 1.92$\pm$0.79 &      &        
%\\
\hline 
$\Omega^-({3\over 2}-)$
& 1.4 to 2.4 &    -           &    -          & 2.25 & 2.25(?) \\
&            &  3.11$\pm$0.13 & 2.31$\pm$1.11 &      &        
\\
&            &  3.18$\pm$0.24 & 3.47$\pm$2.27 & 2.38 & 2.38(?) \\
&            &  3.18$\pm$0.13 & 3.13$\pm$1.50 &      &         
\\
&            &  3.24$\pm$0.24 & 4.32$\pm$2.77 & 2.47 & 2.47(?) \\
&            &  3.24$\pm$0.14 & 3.91$\pm$1.82 &      &         
\\
\hline 
\end{tabular}
\end{table}
%
%%%%%%%%%%%%%%%%%%%%%%%%%%%%%%%%%%%%%%%%%%%%%%%%%%%%%
%%

%%%%%-------------------------------------------------------------------
\section{Conventional QCD sum rules for spin-1/2 decuplet states}
\label{qcd13}

The two sum rules that couple purely to spin-1/2 states are 
from the Dirac structures $\gamma_\mu p_\nu+\gamma_\nu p_\mu$ and 
$\gamma_\mu \gamma_\nu+\frac{1}{3}g_{\mu\nu}$.
Since they cannot be cast into the $(\gamma_0 A + B)$ form in the rest frame,
the parity projection technique cannot be applied to this case.
We have to rely on the conventional sum rule method in which the parities 
are mixed.
The spin-$1/2$ decuplets are in fact excited states. They are
rarely studied in the QCD sum rule method. Here we have an opportunity to 
isolate them as the ground-state poles. For this reason, 
even a conventional analysis is beneficial.

The sum rule from the $\gamma_\mu p_\nu+\gamma_\nu p_\mu$ structure is
\begin{eqnarray}
&  &  c_1\;[1]\; L^{4/27}\; E_2\; M^6
+ c_2\; [b]\; L^{4/27}\; E_0\; M^2
\nonumber \\ & &
+ c_3\; [m_s\,a]\; L^{4/27}\; E_0\; M^2
+ c_4\; [m_s\,m^2_0 a]\; L^{-10/27}
\nonumber \\ & &
+ c_5\; [\kappa_v a^2]\; L^{28/27}
+ c_6\; [m^2_0 a^2]\; L^{14/27}\; {1\over M^2}
\nonumber \\ & &
 = \tilde{\alpha}_{1/2-}^2\; e^{-m^2_{1/2-}/M^2}
+  \tilde{\alpha}_{1/2+}^2\; e^{-m^2_{1/2+}/M^2},
\label{sum33c}
\end{eqnarray}
where the coefficients are given by
\begin{equation}
\begin{array}{|c|c|c|c|c|c|c|}
\hline
         &c_1        &c_2          &c_3             & c_4               &c_5                 &c_6    \\
\hline
\Delta   &{1\over 240}&{5\over 1728}&0               &0                  &{-1\over 18}       &{7\over 216}     \\
\hline 
\Sigma^* &{1\over 240}&{5\over 1728}&{(f_s-4)\over 72}&{(7-4f_s)\over 216}&{-(1+2f_s)\over 54}  &{7(1+2f_s)\over 648} \\
\hline 
\Xi^*    &{1\over 240}&{5\over 1728}&{-(2+f_s)\over 36}&{(7-f_s)\over 36}&-{f_s(2+f_s)\over 54}&{7f_s(1+2f_s)\over 648} \\
\hline 
\Omega^-   &{1\over 240}&{5\over 1728}&{-f_s\over 8}  &{f_s\over 24}    &{-f_s^2\over 18}    &{7f_s^2\over 216}      \\
\hline 
\end{array}
\end{equation}
This sum rule is chiral-even. Since the parity of the lowest state is not known, 
we retain one term for each parity in the RHS, and let the OPE reveal which one is lower.  
The sum rule for $\Delta$ agrees with that given in Ref.~\cite{Derek90}, 
except that the coefficient $c_6$ has the opposite sign.
The sum rules for $\Sigma^*$, $\Xi^*$, and $\Omega^-$ are new, to the best of our knowledge.
Note that the sum rule for $\Omega^-$ reduces to that for $\Delta$ 
if the strange quark is turned off ($m_s=0$ and $f_s=1$), as expected.
Since they are derived separately, this provides a non-trial check of the calculation.
Another check is provided by the fact that $c_5$ and $c_6$ for $\Sigma^*$, $\Xi^*$, and $\Omega^-$ 
coincide with each other in the limit of $f_s=1$, as expected.
The excited state contributions of RHS are modeled using terms
on the OPE side surviving $M^2\rightarrow \infty$ under the assumption
of duality, and are represented by the factors
$E_n(x)=1-e^{-x}\sum_n{x^n/n!}$ with $x=w^2/M^2$
and $w$ an effective continuum threshold.

The sum rule from the 
$ \gamma_\mu\gamma_\nu + {1\over 3}g_{\mu \nu}$ structure, which is chiral-odd, 
is given by
\begin{eqnarray}
&  & c_1\; [ a]\; L^{16/27}\; E_1\; M^4
+c_2\; [m^2_0 a]\; L^{2/27}\; E_0\; M^2
\nonumber \\ & &
+c_3\; [m_s\, b]\; L^{-8/27}\; E_0\; M^2
+c_4\; [a\,b]\; L^{16/27}
\nonumber \\ & &
+c_5\; [m_s\,\kappa_v a^2]\; L^{16/27}
\nonumber \\ & &
= -\tilde{\alpha}_{1/2+}^2 m_{1/2+} e^{-m^2_{1/2+}/M^2}
\nonumber \\ & &
 +\tilde{\alpha}_{1/2-}^2 m_{1/2-}  e^{-m^2_{1/2-}/M^2},
\label{sum33d}
\end{eqnarray}
where the coefficients are given by
\begin{equation}
\begin{array}{|c|c|c|c|c|c|}
\hline
         & c_1                & c_2              & c_3        & c_4                  & c_5  \\
\hline
\Delta   &{-1\over 36        }&{1\over 18}       &  0         &{-5/864}              & 0            \\
\hline
\Sigma^* &{-(2+f_s)\over 108 }&{(2+f_s)\over 54} &{1\over 288}&{-5(2+f_s)\over 2592} &{(f_s-1)\over 18} \\
\hline
\Xi^*    &{-(1+2f_s)\over 108}&{(1+2f_s)\over 54}&{1\over 144}&{-5(1+2f_s)\over 2592}&{f_s(f_s-1)\over 18} \\
\hline
\Omega^- &{-f_s\over 36      }&{f_s\over 18}     &{1\over 96 }&{-5f_s/864}           & 0               \\
\hline
\end{array}
\label{coe33d}
\end{equation}
For $\Delta$, the coefficient $c_4$ is different in both value and sign from that in Ref.~\cite{Derek90}.
The sum rules for $\Sigma^*$, $\Xi^*$, and $\Omega^-$ are new, as far as we know.
Similarly, the sum rule for $\Omega^-$ reduces to that for $\Delta$ 
if the strange quark is turned off.
Another check is provided by the fact that $c_1$, $c_2$, $c_4$ and $c_5$ 
for $\Sigma^*$, $\Xi^*$, and $\Omega^-$ coincide with each other 
in the limit of $f_s=1$.

Now we turn to the analysis of the two sum rules.
First, we note that states of opposite parities on the phenomenological side (RHS) 
are adding up in the chiral-even sum rule, whereas canceling in the chiral-odd sum rule. 
This is a standard feature of baryon sum rules that leads to the general conclusion that 
chiral-odd sum rules perform better than chiral-even sum rules in baryon channels.
Indeed, we found that the chiral-even sum rule in Eq.~(\ref{sum33c}) is very poor.
The leading term is a perturbative contribution, and the sum rule is almost completely saturated 
by the continuum.  No results could be extracted from this sum rule.

\begin{figure*}
\centerline{\psfig{file=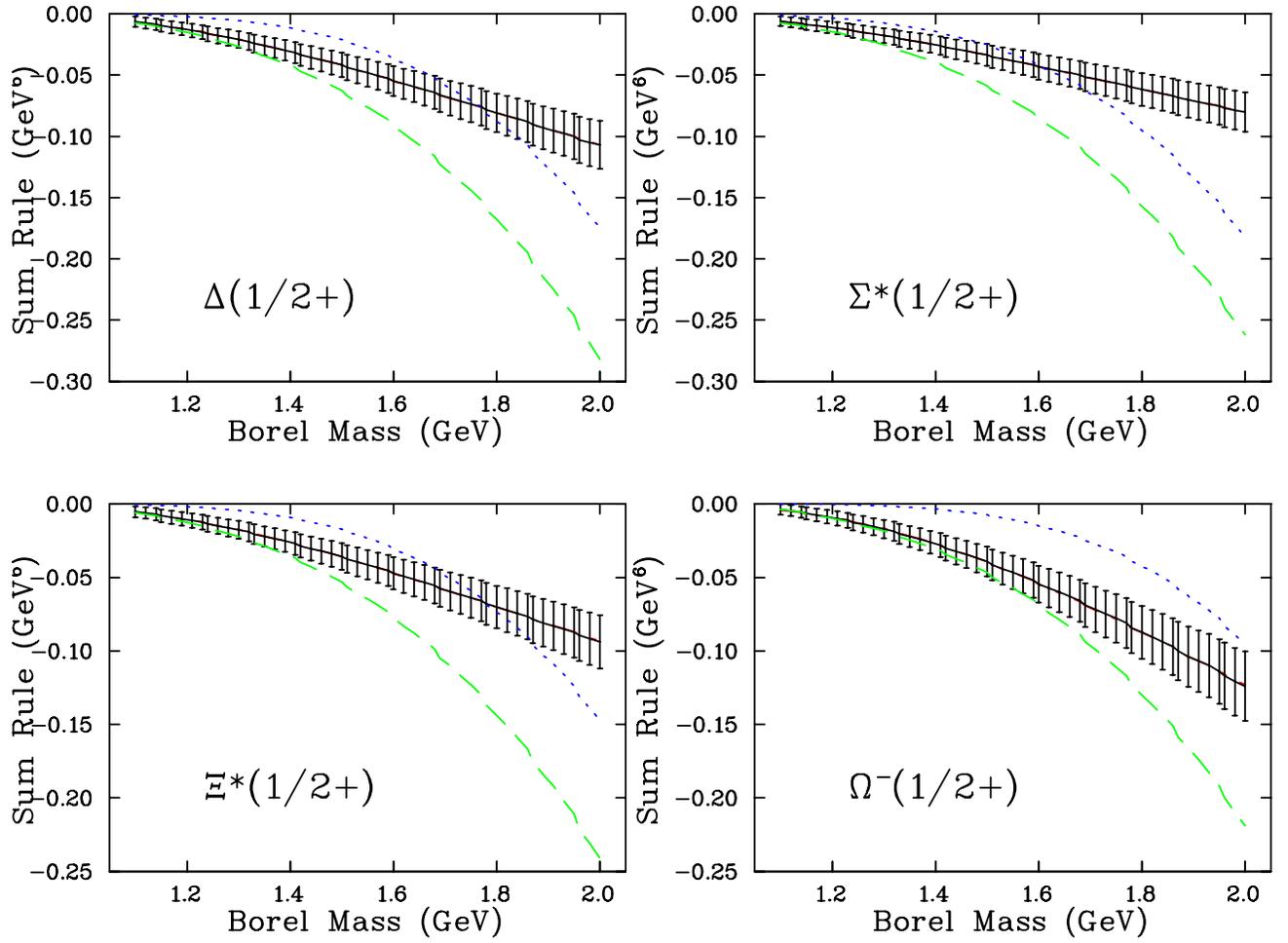,width=5.0in,angle=90}}
\caption{Matching of the sum rules in Eq.~(\protect\ref{sum33d})
for the spin-1/2 decuplet states.
The solid or black line along with the error band generated via Monte-Carlo 
is the LHS (OPE minus continuum).
The dash-dot or red line, which is barely distinguishable from the solid line, 
is the RHS (ground-state pole).
Also plotted are the individual contributions of the OPE (dashed or green line) and 
the continuum (dotted or blue line).}
\label{fit4}
\end{figure*}

On the other hand, the chiral-odd sum rule in Eq.~(\ref{sum33d}) has good convergence.
The leading term contains the non-perturbative quark condensate. 
The sign of this term (in $c_1$) indicates that the sum rule 
is saturated by the positive-parity state:
they have the same negative signs, see Eq.~(\ref{sum33d}) and Eq.~(\ref{coe33d}). 
This is an example of how the parity of the ground-state pole 
is determined in a mixed-parity sum rule.
The dominance of the quark-condensate term is further confirmed in our 
numerical Monte-Carlo analysis.  We are able to perform three-parameter searches. 
It means that the extracted mass, coupling, continuum threshold can be regarded as the 
true predictions of the sum rule.
The results are given in Table~\ref{tab33d} and the matching of the two sides is shown 
in Fig.~\ref{fit4}.
In the observed spectrum, the lowest $\Delta(1/2+)$ is a one-star state at 1.75 GeV, 
followed by a 4-star state at 1.91 GeV. 
Our prediction favors the latter as the lowest state in this channel.
It casts doubts on the existence of the state at 1.75 GeV.
For $\Sigma^*(1/2+)$, the PDG lists a 3-star state at 1.66 GeV, followed by a one-star state 
at 1.77 GeV, then by a 2-star state at 1.88 GeV.
Our result seems to favor the state at 1.88 GeV than the state at 1.66 GeV.
The spin-parity situation in the $\Xi^*(1/2+)$ channel is not clear in the PDG. 
It lists two 3-star states with unknown spin-parity, at 1690 MeV and 1950 MeV.
Our results is in favor of the 1950 MeV state.
In the $\Omega^-$ channel, there are 3 states sitting close to each other 
in the observed spectrum,
at 2250, 2380, and 2470 MeV, whose spin and parity are not clear.
Our prediction of 2.37 GeV is in the middle of this range, 
but the accuracy is not enough to clearly identify with one of them.

%%%%%%%%%%%%%%%%%%%%%%%%%%%%%%%%%%%%%%%%%%%%%%%%%%%%%
%
\begin{table}
\caption{Results for the $1/2+$ branch of the decuplet states 
from the chiral-odd sum rule Eq.~(\ref{sum33d}) in a three-parameter search. 
The errors are derived from 500 Monte-Carlo sets based on the uncertainties assigned 
to the QCD input parameters. 
Two sets of solutions are provided for each case: 
the first one using the default errors in the OPE input parameters, 
while the second using uniform 10\% errors.
The experimental values are taken from the PDG~\protect\cite{pdg04}.}
\label{tab33d}
\begin{tabular}{c|ccccl}
\hline
         & Region & $w$ & $\tilde{\alpha}_{1/2+}^2$ & Mass & Exp.\\
         & (GeV) &  (GeV) & (GeV$^6$) & (GeV) & (GeV)\\ \hline
$\Delta(1/2+)$
& 1.1 to 2.0 & 2.41$\pm$1.08 & 0.18$\pm$0.15  & 2.10$\pm$0.54  & 1.91 \\
&            & 2.52$\pm$0.80 & 0.18$\pm$0.13  & 2.17$\pm$0.33  &     \\
\hline
$\Sigma^*(1/2+)$
& 1.1 to 2.0 & 2.15$\pm$1.06 & 0.13$\pm$0.13  & 1.97$\pm$0.62  & 1.88 \\
&            & 2.29$\pm$0.83 & 0.13$\pm$0.11  & 2.09$\pm$0.37  &     \\
\hline
$\Xi^*(1/2+)$
& 1.1 to 2.0 & 2.42$\pm$1.11 & 0.16$\pm$0.13  & 2.11$\pm$0.54  & 1.95(?) \\
&            & 2.50$\pm$1.50 & 0.16$\pm$0.12  & 2.19$\pm$0.35  &     \\
\hline
$\Omega^{-}(1/2+)$
& 1.1 to 2.0 & 3.02$\pm$1.18 & 0.23$\pm$0.14  & 2.37$\pm$0.41  & 2.38(?) \\
&            & 3.11$\pm$1.13 & 0.23$\pm$0.13  & 2.39$\pm$0.32  &     \\
\hline
\end{tabular}
\end{table}
%
%%%%%%%%%%%%%%%%%%%%%%%%%%%%%%%%%%%%%%%%%%%%%%%%%%%%%

%%%%%----------------------------------------------------------------------
\section{Conclusion}
\label{con}

We have presented a study of the decuplet family using the method of QCD sum rules.
New sum rules are derived for the $3/2+$ and $3/2-$ branches 
using a parity-projection technique. They are more stable and give a 
better determination of the mass spectrum than the conventional sum rules.
The results for the $3/2-$ branch are new as a consequence of the parity separation.
The spin-1/2 sector is investigated using the conventional sum rules method.
New sum rules are derived for members that contain the 
strange quark ($\Sigma^*$, $\Xi^*$, and $\Omega^-$), in addition to a 
careful re-examination of $\Delta$ channel.
The chiral-even sum rules of Eq.~(\ref{sum33c}) are dominated by the continuum. 
No useful information could be extracted from them.
The chiral-odd sum rules of Eq.~(\ref{sum33d}) have good convergence and allow 
the only three-parameter searches in this study. The predicted results 
provide useful information on the $1/2+$ states from a QCD-based standpoint. 
We could not find sum rules that are saturated by the $1/2-$ states.

\begin{acknowledgments}
This work has been supported by DOE under grant number DE-FG02-95ER40907.
\end{acknowledgments}

%%%%--------------------------------------------------------------------------------

\end{document}